\documentclass[aps,prb]{revtex4}
\usepackage{graphicx}

\begin{document}
\pagestyle{empty}
\title{Quantum friction}

\author{A.I.Volokitin$^{1,2}$\footnote{Corresponding author.
\textit{E-mail address}:alevolokitin@yandex.ru}    and B.N.J.Persson$^1$}
 \affiliation{$^1$Institut f\"ur Festk\"orperforschung,
Forschungszentrum J\"ulich, D-52425, Germany} \affiliation{
$^2$Samara State Technical University, 443100 Samara, Russia}

\begin{abstract}
We investigate  the van der Waals friction between graphene and an
amorphous SiO$_2$ substrate.  We find that  due to this friction
the electric current is saturated at a high electric field , in
agreement with experiment. The saturation current depends weakly
on the temperature, which we attribute to the quantum friction
between the graphene carriers and the substrate optical phonons.
We calculate also the frictional drag between two graphene sheets
caused by van der Waals friction, and find that this drag can
induce a  voltage  high enough to be easily measured
experimentally.
\end{abstract}

\maketitle


\vskip 5mm For several decades physicists have been intrigued by
the idea of quantum friction. It has recently been shown that two
bodies moving relative to each other experience a friction due to
quantum fluctuations inside the bodies
\cite{Pendry1997,Volokitin99,Volokitin2008b,VolokitinRMP07}.
 However at present
there is no experimental evidence for or against this effect,
because the predicted friction forces are very small  and
precision measurement of quantum forces are incredibly difficult
with present technology. The existence of quantum friction is
still debated even among theoreticians
\cite{Philbin2009,Pendry2010a,Philbin2010,Pendry2010b,Volokitin2010}.
In this Letter we propose that quantum friction can be observed in
experiments on studying of electrical transport phenomena in
nonsuspended graphene on amorphous SiO$_2$ substrate.

 Graphene, the recently isolated single-layer carbon sheet, consist
 of carbon atoms closely packed
in a flat two-dimensional crystal lattice. The unique electronic
and mechanical properties of graphene\cite{Geim2004, Geim2005} are
being actively explored both theoretically and experimentally
because of its importance for fundamental physics, and for
possible technological applications, in particular for electronics
and sensors \cite{Geim2004,Geim2007}.

Graphene, as  all media, is surrounded  by a fluctuating
electromagnetic field due to the thermal and quantum fluctuations
of the current density.  Outside the bodies this fluctuating
electromagnetic field exists partly in the form of propagating
electromagnetic waves and partly in the form of evanescent  waves.
The theory of the fluctuating electromagnetic field was developed
by Rytov \cite{Rytov53,Rytov67,Rytov89}. A great variety of
phenomena such as Casimir-Lifshitz forces \cite{Lifshitz54},
near-field radiative heat transfer \cite{Polder1971}, noncontact
friction \cite{Volokitin99,VolokitinRMP07,Volokitin2008b}, and the
frictional drag in low-dimensional systems
\cite{Volokitin2001b,Volokitin2008a}, can be described using this
theory.

In this Letter, we investigate the friction force on moving with
drift velocity $v$ charge carriers in graphene due to  interaction
with optical phonons in nearby amorphous SiO$_2$. The dissipated
energy result in heating of the graphene, and is transferred to
the SiO$_2$ substrate via the near-field radiative heat transfer
process and direct phononic coupling. Using the theories of the
van der Waals friction and the near-field radiative heat transfer
we formulate a theory that describes these phenomena and allows us
to predict experimentally measurable effects.  In comparison with
the existing microscopic theories of transport in graphene
\cite{Perebeinos2009,Perebeinos2010}  our theory is macroscopic.
The electromagnetic interaction between graphene and a substrate
is described by the dielectric functions of the materials which
can be accurately determined from theory and experiment.

\begin{figure}
\includegraphics[width=0.45\textwidth]{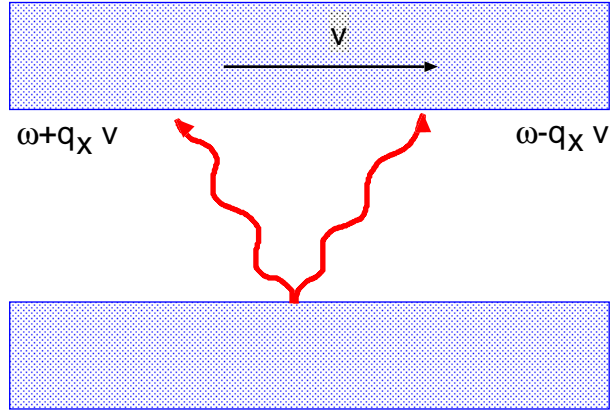}
\caption{\label{Fig1} Two bodies moving relative to each other
will experience van der Waals friction due to Doppler shift of the
electromagnetic waves emitted by them. }
\end{figure}

Assume that a graphene sheet is separated from the substrate by a
sufficiently wide insulator gap, which prevents particles from
tunneling across it. If the charge carriers inside graphene move
with velocity  $v$ relative to the substrate, a frictional stress
will act on them. This frictional stress is related to an
asymmetry of the reflection amplitude along the direction of
motion; see Fig. \ref{Fig1}.
If the substrate emits radiation, then in the rest reference frame
of charge carriers in graphene these waves are Doppler shifted
which will result in different reflection amplitudes. The same is
true for radiation emitted by moving charge carriers in graphene.
The exchange of ``Doppler-shifted-photons'' will result in
momentum and energy transfer between graphene and substrate, which
is the origin of the van der Waals friction.

 Let us consider graphene and a substrate with
flat parallel surfaces at separation $d\ll \lambda_T=c\hbar/k_BT$.
Assume that the free charge carriers in graphene move  with the
velocity $v\ll c$  ($c$ is the light velocity) relative to the
substrate. According to Ref.
\cite{Volokitin99,VolokitinRMP07,Volokitin2008b} the frictional
stress $F_x$ acting on charge carriers in graphene, and the
radiative heat flux $S_z$ across the surface of substrate, both
mediated by a fluctuating electromagnetic field, are determined by

\[
F_x =\frac \hbar {\pi ^3}\int_{0 }^\infty dq_y\int_0^\infty
dq_xq_xe^{-2qd}\Bigg \{ \int_0^\infty d\omega \Bigg(
\frac{\mathrm{Im}R_{d}(\omega)\mathrm{Im}R_{g}(\omega^+) }{\mid
1-e^{-2 q d}R_{d}(\omega)R_{g}(\omega^+)\mid ^2}\times
\]
\[
 [n_d(\omega )-n_g(\omega^+)]+\frac{\mathrm{Im}R_{d}(\omega^+)\mathrm{Im}R_{g}(\omega^) }{\mid
1-e^{-2 q d}R_{d}(\omega^+)R_{g}(\omega)\mid ^2}[n_g(\omega
)-n_d(\omega^+)]\Bigg )+
\]
\begin{equation}
 \int_0^{q_xv}d\omega \frac{\mathrm{Im}R_{1}(\omega)\mathrm{Im}
R_{2}(\omega^-)} {\mid 1-e^{-2qd}R_{1}(\omega)R_{2}(\omega^-)\mid
^2} [n_2(\omega^-)-n_1(\omega)] \Bigg \}, \label{parallel2}
\end{equation}
\[
S_z =\frac \hbar {\pi ^3}\int_{0 }^\infty dq_y\int_0^\infty
dq_xe^{-2qd}\Bigg \{ \int_0^\infty d\omega \Bigg(- \frac{\omega
\mathrm{Im}R_{1}(\omega)\mathrm{Im}R_{2}(\omega^+) }{\mid 1-e^{-2
q d}R_{1}(\omega)R_{2}(\omega^+)\mid ^2}\times
\]
\[
 [n_1(\omega )-n_2(\omega^+)]+\frac{\omega^+\mathrm{Im}R_{d}(\omega^+)\mathrm{Im}R_{g}(\omega) }{\mid
1-e^{-2 q d}R_{1}(\omega^+)R_{2}(\omega)\mid ^2}[n_2(\omega
)-n_1(\omega^+)]\Bigg )+
\]
\begin{equation}
 \int_0^{q_xv}d\omega \frac{\omega \mathrm{Im}R_{1}(\omega)\mathrm{Im}
R_{2}(\omega^-)} {\mid 1-e^{-2qd}R_{1}(\omega)R_{2}(\omega^-)\mid
^2} [n_2(\omega^-)-n_1(\omega)] \Bigg \}, \label{parallel2}
\end{equation}
where  $n_i(\omega )=[\exp (\hbar \omega /k_BT_i-1]^{-1}$
($i=g,d$), $T_{g(d)}$ is the temperature of graphene (substrate),
 $R_{i}$  is the reflection amplitude for
surface $i$ for $p$ -polarized electromagnetic waves, and
$\omega^{\pm}=\omega \pm q_xv$.  The reflection amplitude for
graphene (substrate)  is determined by \cite{Volokitin2001b}
\begin{equation}
R_{g(d)}=\frac{\epsilon _{g(d)}-1}{\epsilon _{g(d)}+1},
 \label{refcoef}
\end{equation}
where $\epsilon _{g(d)}$ is the dielectric function for graphene
(substrate).

In the study below we used the dielectric function of graphene, which was
calculated recently within the random-phase approximation (RPA)
\cite{Wunsch2006,Hwang2007}. The small (and constant) value of
the graphene Wigner-Seitz radius $r_s$ indicates that it is a weakly
interacting system for all carries densities, making the RPA an
excellent approximation for graphene (RPA is asymptotically
exact in the $r_s\ll1$ limit).  The dielectric function
is an analytical function in the upper half-space
of the complex $\omega$-plane:
\begin{equation}
\epsilon_g(\omega,q)=1+\frac{4k_Fe^2}{\hbar
v_Fq}-\frac{e^2q}{2\hbar \sqrt{\omega^2-v_F^2q^2}}\Bigg \{G\Bigg
(\frac{\omega+2v_Fk_F}{v_Fq}\Bigg )- G\Bigg
(\frac{\omega-2v_Fk_F}{v_Fq}\Bigg )-i\pi \Bigg \},
\end{equation}
where
\begin{equation}
G(x)=x\sqrt{1-x^2} - \ln(x+\sqrt{1-x^2}),
\end{equation}
where the Fermi wave vector $k_F=(\pi n)^{1/2}$, $n$ is the
concentration of charge carriers, the Fermi energy
$\epsilon_F=\gamma k_F=\hbar v_Fk_F$, $\gamma=\hbar v_F\approx
6.5$ eV\AA, and $v_F$ is the Fermi velocity. The dielectric
function of amorphous SiO$_2$ can be described using an oscillator
model\cite{Chen2007}.

The equilibrium or steady state temperature can be obtained from the condition that
the heat power generated by friction must be equal to the heat transfer
across the substrate surface
\begin{equation}
F_x(T_d,T_g)v=S_z(T_d,T_g)+\alpha_{ph}(T_g-T_d), \label{eqtemp}
\end{equation}
where the second term in Eq. (\ref{eqtemp}) takes into account the
heat transfer through direct phononic coupling; $\alpha_{ph}$
is the thermal contact conductance due to phononic coupling.

\begin{figure}
\includegraphics[width=0.70\textwidth]{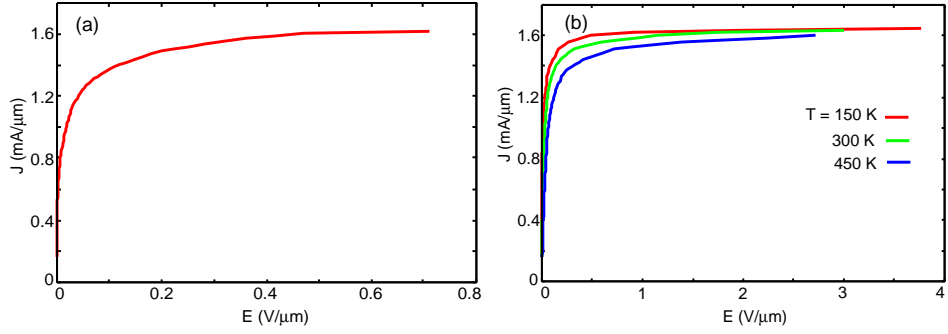}
\caption{\label{Fig2} The role of the interaction between phonon
polaritons in SiO$_2$ and free carriers in graphene for graphene
field-effect transistor transport. The separation between graphene
and SiO$_2$ is $d=3.5$\AA.  (a) Current density-electric field
dependence at $T=0$ K, $n=10^{12}$cm$^{-12}$. (b) The same as in
(a) but for different temperatures.}
\end{figure}

   Figures. \ref{Fig2}a and \ref{Fig2}b show the dependence of the current
density on the electric field at $n=10^{12}$cm$^{-2}$, and for
different temperatures.  In obtaining these curves we have used
that $J=nev$ and $neE=F_x$, where $J$ and $E$ are current density
and electric field, respectively.  Note that the current density
saturate at $E\sim 0.5-2.0$V/$\ \mu$m, which is in agreement  with
experiment \cite{Freitag2009}. The saturation velocity can be
extracted from the $I-E$ characteristics using
$J_{sat}=nev_{sat}$, where 1.6 mA/$\mu$m is the saturated current
density, and with the charge density concentration
$n=10^{12}$cm$^{-2}$: $v_{sat}\approx 10^6$m/s. Fig. \ref{Fig2}a
was calculated at $T_d=0$ K. At zero temperature the van der Waals
friction is due to quantum fluctuations of charge density, and is
determined by the second term in Eq. (\ref{parallel2})
\cite{Pendry1997,Volokitin99,VolokitinRMP07,Volokitin2008b}
\begin{equation}
F_x(T_d=T_g=0) =-\frac\hbar {\pi ^3}\int_{0 }^\infty
dq_y\int_0^\infty dq_x\int_0^{q_xv}d\omega q_xe^{-2qd}
\frac{\mathrm{Im}R_{d}(\omega)\mathrm{Im} R_{g}(\omega^-)} {\mid
1-e^{-2qd}R_{d}(\omega)R_{g}(\omega^-)\mid ^2}
\end{equation}
The existence of quantum friction is still debated in the
literature
\cite{Philbin2009,Pendry2010a,Philbin2010,Pendry2010b,Volokitin2010}.
The van der Waals friction can be studied in noncontact
experiments, and in frictional drag experiments
\cite{VolokitinRMP07}. In both these experiments the solids are
separated by a potential barrier thick enough to prevent electrons
or other particle with a finite rest mass from tunneling across
it, but allowing the interaction via the long-range
electromagnetic field, which is always present in the gap between
bodies. In noncontact friction experiments the damping of
cantilever vibrations is typically measured, while in frictional
drag experiments a current density is induced in one medium. The
friction between the moving charge carriers and nearby medium
gives rise to a change of $I-E$ characteristics, which can be
measured.

The friction force acting on the charge carriers in graphene for
high electric field is determined by the interaction with the
optical phonons of the graphene, and with the optical phonons of
the substrate. The frequency of optical phonons in graphene is a
factor 4 larger than for the optical phonon in SiO$_2$. Thus, one
can expect that for graphene on SiO$_2$ the high-field $I-E$
characteristics will be determined by excitations of optical
phonons in SiO$_2$. According to the theory of the van der Waals
friction \cite{VolokitinRMP07}, the quantum friction, which exists
even at zero temperature, is determined by the creation of
excitations in each of the interacting media, the frequencies of
which are connected by $vq_x=\omega_1 + \omega_2$. The relevant
excitations in graphene are the electron-hole pairs whose
frequencies begin from zero, while for SiO$_2$ the frequency of
surface phonon polaritons $\omega_{ph} \approx 60$meV ($9\cdot
10^{13}$s$^{-1}$). The characteristic wave vector of graphene is
determined by Fermi wave vector $k_F$. Thus the friction force is
strongly enhanced when $v>v_{sat}=\omega_{ph}/k_F \sim 10^6$m/s,
in  accordance with numerical calculations. Thus  measurements of
the current density-electric field relation of graphene adsorbed
on SiO$_2$ give the possibility to detect quantum friction.

An alternative method of studying of the van der Waals friction
consists in driving an electric current in one metallic layer and
studying of the effect of the frictional drag on the electrons in
a second (parallel) metallic layer.
\begin{figure}
\includegraphics[width=0.70\textwidth]{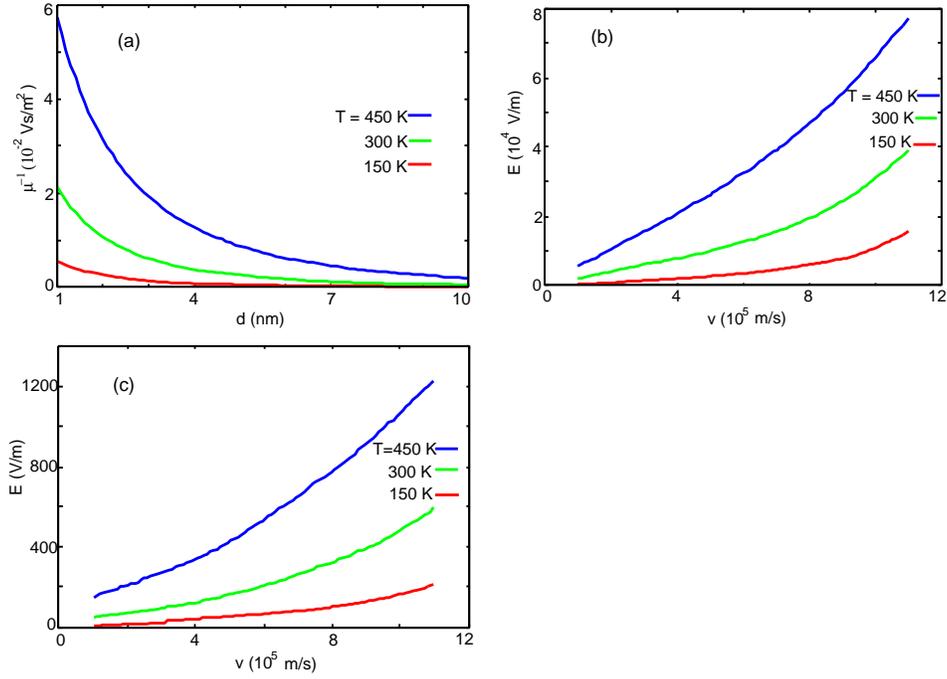}
\caption{\label{Fig4} Frictional drag between two graphene sheets
at the carrier concentration $n=10^{12}$cm$^{-2}$.  (a) Dependence
of friction coefficient per unit charge, $\mu^{-1}=\Gamma/ne$, on
the separation between graphene sheets $d$. (b) Dependence of
electric field induced in graphene on drift velocity of charge
carriers in other graphene sheet at the layer separation $d=1$ nm.
(c) The same as in (b) but at $d=10$nm.}
\end{figure}
Such experiments were first suggested by Pogrebinskii
\cite{Pogrebinskii} and Price \cite{Price}, and were performed for
2D-quantum wells \cite{Gramila1,Sivan}.

Similar to 2D-quantum wells in semiconductors, frictional drag
experiments can be performed (even more easily) between graphene
sheets. Such experiments can be performed in a vacuum where the
contribution from the phonon exchange can be excluded. To exclude
noise (due to presence of dielectric) the frictional drag
experiments between quantum wells were performed at very low
temperature ($T\approx 3 \ {\rm K}$)\cite{Gramila1}. For suspended
graphene sheets there is no such problem and the experiment can be
performed at room temperature. In addition, 2D-quantum wells in
semiconductors have very low Fermi energy $\epsilon_F \approx
4.8\times 10^{-3}$eV \cite{Gramila1}. Thus electrons in these
quantum wells are degenerate only for very low temperatures
$T<T_F=57$ K. For graphene the Fermi energy $\epsilon_F=0.11$eV at
$n=10^{12}$cm$^{-2}$, and the electron gas remains degenerate for
$T<1335$ K.

At small velocities the electric field induced by frictional drag
depends linearly on the velocity, $E=(\Gamma/ne)v = \mu^{-1}v$,
where $\mu$ is the low-field mobility. For $\hbar \omega \ll
\epsilon_F$ and $q\ll k_F$ the reflection amplitude for graphene
is given by the same expression as for a 2D-quantum well
\cite{VolokitinRMP07} and from  Eq. (\ref{refcoef}) we get
\begin{equation}
\Gamma =
0.01878\frac{\hbar}{d^4}\left(\frac{k_BT}{k_Fe^2}\right)^2
\end{equation}

Figure \ref{Fig4}a shows the dependence of the friction coefficient
(per unit charge) $\mu^{-1}$ on the separation $d$ between the
sheets. For example, $E=5\times 10^{-4}v$ for $T=300$ K and $d=10$
nm. For a graphene sheet of length $1\ {\rm \mu m}$, and with
$v=100$m/s this electric field will induce the voltage $V=10$ nV.
Figures \ref{Fig4}b  and \ref{Fig4}c show the induced electric
field-velocity relation for high velocity, with $d=1$nm (b) and
$d=10$nm (c).

 \textit{Concluding remarks}.--We have used theories of the van der Waals friction and near-field
 radiative heat transfer to study
transport in graphene due to the interaction with
phonon-polaritons in an (amorphous) SiO$_2$ substrate. High-field
transport exhibit a weak temperature dependence, which can be
considered as manifestation of quantum fluctuations. Thus the
study of transport properties in graphene gives the possibility of
detecting quantum friction, the existence of which is still debated
in the literature. We have calculated the frictional drag between
graphene sheets mediated by the van der Waals friction, and found
that it can induce large enough voltage to be easily measured.

 \vskip 0.5cm
\textbf{Acknowledgment}

A.I.V acknowledges financial support from the Russian Foundation
for Basic Research (Grant N -10-02-00297-à) and ESF within activity
``New Trends and Applications of the Casimir Effect''.

\end{document}